\documentclass[12pt]{article}
\usepackage{amsmath}
\usepackage{graphicx}
\usepackage{verbatim}
\usepackage{graphicx}
\usepackage{color}

\newcommand{\etal}{\emph{et}$\,$\emph{al.} }

\newcommand{\romd}{{\operatorname{d}}}

\newcommand{\romG}{{\operatorname{G}}}

\newcommand{\VECf}{{\bf f}}

\newcommand{\VECm}{{\bf m}}
\newcommand{\VECn}{{\bf n}}

\newcommand{\VECt}{{\bf t}}
\newcommand{\VECu}{{\bf u}}
\newcommand{\VECz}{{\bf z}}

\newcommand{\VECF}{{\bf F}}
\newcommand{\VECJ}{{\bf J}}
\newcommand{\VECM}{{\bf M}}
\newcommand{\VECX}{{\bf X}}

\begin{document}

\renewcommand{\thefootnote}{\fnsymbol{footnote}}

\sf
\begin{center}
   \vskip 2em
     {\LARGE \sf How paper folds: bending with local constraints}
  \vskip 3em
  {\large \sf
Jemal Guven${}^{(1)}$ and Martin Michael
M\"uller$^{(2),}$\footnote{Current address: Laboratoire de
Physique Statistique, Ecole Normale Sup\'erieure, 24, rue Lhomond,
75231 Paris Cedex 05, FRANCE}\addtocounter{footnote}{-1}
 \\[2em]}
\em{ ${}^{(1)}$ Instituto de Ciencias Nucleares,
 Universidad Nacional Aut\'onoma de M\'exico\\
 Apdo. Postal 70-543, 04510 M\'exico, D.F., MEXICO\\[1em]
${}^{(2)}$ Max-Planck-Institut f\"ur Polymerforschung\\
Ackermannweg 10, 55128 Mainz, GERMANY\\[1em]
}

\end{center}
 \vskip 1em

 \renewcommand{\thefootnote}{\arabic{footnote}}


\begin{center}
\today
\end{center}

\begin{abstract}
\sf A variational framework is introduced to describe how a
surface bends when it is subject to local constraints on its
geometry. This framework is applied to describe the  patterns of a
folded sheet of paper. The unstretchability of paper implies a
constraint on the surface metric; bending is penalized by an
energy quadratic in mean curvature. The local Lagrange multipliers
enforcing the constraint are identified with a conserved
tangential stress that couples to the extrinsic curvature of the
sheet. The framework is illustrated by examining the deformation
of a flat sheet into a generalized cone.
\end{abstract}


\section{\Large \sf Introduction\label{sec:introduction}}

There is a striking similarity in the patterns we observe in a
crumpled ball of paper, a crushed aluminium can or, for that
matter, the creases that develop on the back of one's shirt when
one's back has been up against the wall a little too long. This is
not an accident.

\vskip1pc\noindent The relevant property that these materials
share is that they offer resistance to tangential strain. This is
captured geometrically in the statement that the metric, which
determines distances between points on the surface (and thus is a
quantity intrinsic to it), does not change under deformation: the
only permissible deformations are isometries
\cite{HCV,Duncans,Fuchs,FuchsBook}. Not only must the surface area
remain fixed, right angles must remain right on the surface.

\vskip1pc\noindent
In general, the number of ways that a surface can bend without
stretching is limited. In particular, if a flat sheet is bent in
more than one direction it must stretch somewhere. This is a
direct consequence of the geometrical fact--Gauss's inspired
insight--that the product of the curvatures at any point on the
surface depends only on the metric there \cite{Struik}: two
quantities that are defined extrinsically are constrained by the
intrinsic geometry. In particular, if the surface is flat to start
with, so that one or both of its curvatures vanishes, it must
remain flat. However, the number of ways a surface can be flat is
itself limited. It is either part of a cylinder, a cone or, more
generally, a tangent developable surface (think Frank Gehry
\cite{Gehry}), ruled by straight lines \cite{HCV,Struik,Gray}. Any
deformation of the surface is constrained locally by the fact that
it must coincide with one of them.

\vskip1pc\noindent The physically deformed surface is, of course,
rarely any single one of these elementary flat surfaces. This is
because the forces applied to the surface usually oblige it to
fold along more than one direction somewhere. It turns out,
however, that the most energy effective way to adjust to such
forces is by confining the regions where stretching occurs within
a series of sharp peaks and ridges \cite{Witten,benpom}. A tangent
plane cannot be drawn at these singular points; the curvatures
will generally diverge. The complete deformed surface will form a
quilt of irregularly shaped flat patches meeting along the
boundaries that these points provide. As described by Cerda and
Mahadevan they provide the hinges about which the flat patches may
flap \cite{PRS}. Once such singular points are established,
however, irreversible damage will be done to the surface: iron out
the creases we may but the paper will remain scored, the can
scarred and, perhaps this is stretching the analogy, shirts do
become threadbare.

\vskip1pc\noindent A theory which describes the detailed internal
structure of this network of peaks and ridges will be complicated
and it will depend on the detailed material properties of paper.
An approximation, reliable for small strains, is provided by the
F{\"o}ppl-von Karman equations \cite{LL}. The length scale of
these regions is on the order of the thickness of the sheet;
viewed from outside, however, this internal structure is
irrelevant: peaks become points and ridges can be treated as
curves along which boundary conditions are set on the flat
surface; the only surviving material parameter in the physical
description of the sheet is a modulus characterizing its
stiffness. The important point is that on these longer length
scales the only relevant degrees of freedom are the geometrical
ones; how the surface bends becomes an essentially geometrical
problem, a fact that tends to get lost when it is treated from a
continuum mechanical point of view. In this paper we would like to
explore this geometrical approach to the problem.

\vskip1pc\noindent The bending energy of a surface is  deceptively
simple; modulo a topological energy there is a single quadratic in
curvature associated with a surface \cite{Willmore,Helfrich}.
However, determining the  bending patterns of a surface when
bending alone is penalized provides a subtle non-linear problem in
itself and there are few useful analytical results. The problem we
are interested in addressing has an added complication: there are
local geometrical constraints on the surface; the metric is fixed.

\vskip1pc\noindent When the constraints are global it is
straightforward to enforce them in the minimization of the energy
by introducing Lagrange multipliers. If the enclosed volume and
the area are prescribed, these multipliers are identified with the
equilibrium pressure and tension: they are global. This is a
situation that is familiar (in the context of fluid membranes see,
for example, \cite{Seifert}). With local constraints these
multipliers get replaced by local fields with their own
conservation laws reflecting the constraint.

\vskip1pc\noindent In this paper we will set up a variational
framework to describe the equilibrium shapes assumed by a surface
subject to local constraints. In equilibrium, we will show that
the multipliers are identified with a conserved tangential surface
stress. The constraints fixing the metric thus set up tangential
stresses within the sheet. These add to the stresses associated
with bending. They couple linearly to the extrinsic curvature in
the shape equation describing the equilibrium configuration.

\vskip1pc\noindent We will illustrate the framework by examining
the deformation of a flat sheet into a generalized cone
\cite{benpom}. This is the simplest flat deformation of a plane
exhibiting a localized distribution of energy. Our focus will be
on the stresses set up in the cone rather than the geometrical
details of the configuration itself. We will examine in detail how
the different contributions to the stress conspire to transmit
force and torque along closed curves on the cone. In particular,
we will show how our framework can be applied to two physically
interesting setups that have been discussed in the literature. The
first of these considers the depression of a circular sheet into a
circular frame by application of a point force to its center
\cite{PRS}. Neither the sheet nor the frame needs to be circular.
We will show that the consistent coupling to curvature is enough
to completely determine the radial behavior of the stress tangent
to a sphere centered on the apex of the cone. The radial behavior
of the remaining projections of the stress tensor is then
completely determined by the conservation law. We will show that
additional off-diagonal stresses, that have not been studied
previously, are consistent with the conical geometry. The second
example considers the `draping' patterns assumed by a sheet
supported  at a point that falls under the influence of gravity
\cite{drape}. Additional stress fields are now needed to
counteract the effects of gravity which would tend to {\it
unflatten} the disc. We show, however, that they are also
completely determined by the consistent coupling to curvature.
Remarkably, the shape equation we obtain assumes a universal form,
with or without gravity; the physical details of the problem enter
the equation only through the stress tangent to the sphere
associated with the constraint. In particular, with gravity, this
equation differs from its (technically inaccurate) counterpart
obtained by pre-averaging over the radial direction.


\section{\Large \sf Bending a constrained sheet\label{sec:sheetbending}}

\vskip1pc\noindent We are interested in determining the
configurations that minimize the bending energy of an initially
flat unstretchable  surface under the influence of some set of
external forces or constraints. This might be gravity, forces that
act locally or the confinement of the surface within a fixed
volume. We will first examine the problem without the
complications introduced by gravity.

\vskip1pc\noindent The bending energy associated with the
configuration ${\bf X}$ is given by
\begin{equation}
 H [{\bf X}]  = \frac{1}{2}\int dA\, K^2 \,,
\label{eq:HB}
\end{equation}
where $K = C_1 + C_2$ is the sum of the principal curvatures
(twice the mean curvature), and $dA$ is the area element induced
on the surface (our notation is summarized in \cite{notation}). We
have set the rigidity modulus to be one.



\vskip1pc\noindent
Bending must be an isometry: not only does the geometry of the sheet
remain flat almost everywhere when it is bent, it also resists
shear and stretching. Thus it is not sufficient to demand that the
Gaussian curvature vanishes; one needs to fix the metric itself.
The bending energy $H[{\bf X}]$ defined by Eq.~(\ref{eq:HB}) needs
to be minimized subject to this constraint. The framework will
also need to accommodate discontinuities and singularities which
arise due to global obstructions on this requirement.

\vskip1pc\noindent Distances on the surface are described by the
metric, $g_{ab}= {\bf e}_a\cdot {\bf e}_b$, where ${\bf e}_a
=\partial_a {\bf X}$, $a=1,2$, are the two tangent vectors adopted
to the particular parametrization of the surface. In a sheet of
paper the metric is fixed, $g_{ab}^{(0)}$ say. We thus construct a
functional to reflect this constraint \begin{equation} H_C[{\bf
X}]= H[{\bf X}] - \frac{1}{2}\,\int dA\, T^{ab}(u^a) \, (g_{ab}-
g_{ab}^{(0)}) \,. \label{eq:Hcdef} \end{equation} Here
$T^{ab}(u^a)$ is a set of local Lagrange multipliers implementing
the  constraint on the metric. Both $g_{ab}$ and $g_{ab}^{(0)}$ as
well as the  multipliers $T^{ab}$ form tensors, so that the trace
appearing in Eq.~(\ref{eq:Hcdef}) is a scalar. Thus, by
construction, the functional $H_C$ is independent of the
parametrization.



\vskip1pc\noindent With the constraint in place, it is possible to
introduce a deformation ${\bf X}\to {\bf X}+ \delta {\bf X}$ in the
shape without needing to worry if the metric is behaving.

\vskip1pc\noindent In the absence of external sources of stress, the
constrained equilibrium may be expressed as the conservation law
\cite{stress}
\begin{equation}
\nabla_a {\bf f}^a =0\,, \label{eq:fcons}
\end{equation}
where the stress tensor ${\bf f}^a$ is a sum of two terms:
\begin{equation}
{\bf f}^a = {\bf f}^a_0 + T^{ab}\,{\bf e}_b \,. \label{eq:stressdef}
\end{equation}
The bending contribution is given by (for our notation see
\cite{notation})
\begin{equation}
{\bf f}^a_0=  K (K^{ab}- \frac{1}{2} g^{ab} K)\, {\bf e}_b
-(\nabla^a K)\, {\bf n}\,. \label{eq:stressdef0}
\end{equation}
There
is a particularly direct way to derive these equations.  It involves a natural
extension of an approach to the equilibrium of fluid membranes
using a set of auxiliary variables introduced by one of the
authors \cite{auxil}.

\vskip1pc\noindent The constraint adds a tangential stress proportional to the
multiplier $T^{ab}$; tension will be introduced in the surface
whenever its action is antagonistic to bending. This tension will
generally be inhomogeneous and anisotropic.

\vskip1pc\noindent The `shape equation' is given by the
normal projection of the conservation law~(\ref{eq:fcons}):
\begin{equation}
{\cal E}= {\cal E}_0 - K_{ab} T^{ab} =0\,,
 \label{eq:elch}
\end{equation}
where
\begin{equation}
{\cal E}_0 =-\, \nabla^2 K + \frac{1}{2} K (K^2 - 2 K_{ab} K^{ab}
) \,.
 \label{eq:e0}
\end{equation}
Here $\nabla^2$ is the Laplacian on the surface constructed with
the induced metric. The stress $T^{ab}$ couples linearly to the
extrinsic curvature in Eq.~(\ref{eq:elch}).

\vskip1pc\noindent The tangential component of the conservation law
(\ref{eq:fcons}) is the statement that $T^{ab}$ itself is conserved,
\begin{equation}
\nabla_a T^{ab}=0\,. \label{eq:Tcons0}
 \end{equation}
The bending energy does not enter this equation. It is an
intrinsic statement involving the intrinsic geometry through the
covariant derivative. We note that $T^{ab}$ involves three degrees
of freedom; this is the correct number of components to pin-down
the three components of the metric.  From a field theoretical
point of view it is also possible to view the two constraints
provided by Eq.~(\ref{eq:Tcons0}) abstractly as a requirement for
consistency on the $T^{ab}$ implied by reparametrization
invariance \cite{stress}. Note that the tangential part of ${\bf
f}_0$ defined by Eq.~(\ref{eq:stressdef0}), unlike $T^{ab}$, is
not conserved.

\vskip1pc\noindent It is worth pointing out that this variational
approach to bending with a local constraint is more direct than
its continuum mechanical counterpart. One does not need to invoke
any adapted system of coordinates. One also sidesteps the need to
introduce constitutive relations; once the energy function is
given, the relationship between stress and strain follows.

\vskip1pc\noindent One now needs to solve the coupled system of
equations (\ref{eq:elch}) and (\ref{eq:Tcons0}) subject to
appropriate boundary conditions. Eq.~(\ref{eq:elch}) will
pick up a source if external forces are operating on the sheet. In addition,
the geometry will typically display geometrical singularities where Eq.~(\ref{eq:Tcons0})
will pick up a distributional source. We will apply this framework to the conical
geometry in the next section.

\vskip1pc\noindent It is also instructive to examine how this setup would be different
had we fixed only the Gaussian curvature. This is addressed in
appendix~\ref{app:Gaussiancurvature}. The relevant extension of our framework to
handle the situation in which the initial surface is not flat is a straightforward
generalization.


\section{\Large \sf Folding a cone\label{sec:conefolding}}

\subsection{\Large \sf Geometry of a Developable
Cone\label{sec:conegeometry}}

We begin with a brief review of the conical geometry described by
Ben Amar and Pomeau \cite{benpom}. For further details see
\cite {PRS}.

\begin{figure}
  \begin{center}
  \includegraphics[scale=1.3]{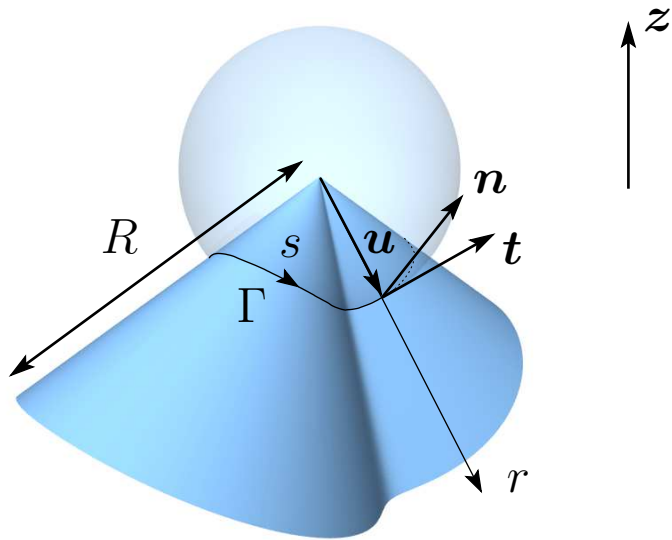}
  \caption{The cone geometry. A closed curve $\Gamma (s)$ on the unit sphere
  is sufficient to describe the surface completely.}
  \label{fig:cone}
  \end{center}
\end{figure}

 \vskip1pc\noindent A cone possesses a natural parametric representation
 in terms of a closed curve $\Gamma: s\to {\bf u}(s)$ on
the unit sphere. Let $s$ be arc-length along this curve and
$r$ be the distance from the origin along the ray pointing in the
direction ${\bf u}(s)$  (see Fig.~\ref{fig:cone}). The image of
the mapping
\begin{equation}
(r,s)\to  {\bf X}(r,s) = r {\bf u}(s)
\end{equation}
describes a cone.
We will suppose that this cone is the result of an isometric
deformation of a planar disc (not necessarily circular) of radius $R(s)$. The length of
$\Gamma$ is then fixed at $L=2\pi$. This means that the closed
curve which represents the cone lives always within a single
hemisphere. A flat planar disc is represented by a great circle.
These are the only circular configurations consistent with the
constraint on the length. If the disc is circular of radius $R$ and
the apex coincides with its center, the maximum value of $r$ is given by $R$.

\vskip1pc\noindent The tangent vectors to the cone adapted to this
parametrization are ${\bf u}$ itself and ${\bf t}={\bf u}'$, where
prime denotes a derivative with respect to $s$. Note that ${\bf
u}\cdot {\bf u}'=0$ so that the induced metric on the surface is
specified by the flat line element
\begin{equation}
dl^2 = dr^2 +  r^2 ds^2\,.
\label{eq:line}
\end{equation}
Let ${\bf n}={\bf u}\times {\bf t}$ denote the normal to the surface.
The extrinsic curvature tensor is given by
\begin{equation} K_{ab}=  r \left( \begin{array}{cc}
  0 & 0\\
  0 & \kappa
  \end{array}\right)
\,,\label{eq:Kab}
\end{equation} where
\begin{equation}
\kappa=  -{\bf n}\cdot  {\bf t}' =- ({\bf u}\times {\bf t})\cdot
{\bf t}' =- {\bf u}\cdot ({\bf t}\times {\bf t}') \,.
\end{equation}
The flat direction is along ${\bf u}$.  The Gaussian curvature
$K_{\romG}=\det{(K_{a}^{b})}$ vanishes.
The curvature $K= \kappa/r$
diverges at the apex of the cone. This translates into a logarithmic
divergence in the bending energy at this point.

\vskip1pc\noindent The three vectors ${\bf u}, {\bf t}$ and ${\bf
n}$ form a right-handed basis. Given $\kappa$, the curve can be
reconstructed using the structure equations
\begin{subequations}
\begin{eqnarray}
{\bf u}' & = & {\bf t} \label{eq:tpu}
\\
{\bf t}' & = & -\kappa {\bf n} - {\bf u} \label{eq:tpt}
\\
{\bf n}' & = & \kappa\,{\bf t} \label{eq:tpn} \,.
\end{eqnarray}\label{eq:tp}
\end{subequations}
\vskip1pc\noindent Note that the vectors ${\bf u}$ and ${\bf n}$
are in general not equal to the normal and binormal vector of
the curve \cite{Gray}. Thus $\kappa$ should not be mistaken for
the Frenet curvature, $\kappa_F$. It is simple, however, to show
that the two are related by $\kappa_F^2 = \kappa^2 +1$.


\subsection{\Large \sf The cone as a constrained
minimum\label{subsec:coneconstrained}}

\vskip1pc\noindent We have seen that cones can be identified with
curves on a unit sphere. As described in
appendix~\ref{app:auxvariables}, by integrating over the radial
direction $H$ becomes a functional of curves on this sphere. As
such, one could vary the Hamiltonian with respect to these curves
to obtain an Euler-Lagrange equation which is satisfied by the
curves which minimize the energy.
This curve is then identified with the equilibrium shape of the
conical surface. While this approach yields the correct shape
equation in this case, it does not always provide the correct
answer. As we will see, variation of the Hamiltonian and
restriction of the geometry (integrating out the radial dependence
in the conical geometry) do \emph{not} commute in general even
when the disc is circular. Even in this case, however, it does not
possess the scope to provide a complete analysis of the problem.

\vskip1pc\noindent We will instead apply our general framework
to study
the conical equilibrium. We first confirm that cones are in fact
equilibrium geometries. It is straightforward to show that
\begin{equation}
{\cal E}_0 =  -\, \nabla^2 K + {1\over 2} K (K^2 - 2 K_{ab} K^{ab}
)   = - {1\over r^3} \, \left(\kappa '' + {1\over 2} \, \kappa^3 +
\kappa\right)\,.
\end{equation}
Note the appearance of the term linear in $\kappa$ originating in
the radial part of the Laplacian. The shape equation thus reads
\begin{equation}
- {1\over r^3} \left(\kappa '' + {1\over 2} \, \kappa^3 + \kappa
\right) - \kappa\, r \, T^{ss} =0\,. \label{eq:ELkappaT}
\end{equation}
Only the stress tangential to the generating curve on the unit sphere contributes.

\vskip1pc\noindent The shape equation describing the conical
equilibrium must hold for all $r$. Indeed, consistency requires
that $r^4 T^{ss}$ depends only on $s$. Let us write $r^4 T^{ss} =
- C_{\|}(s)$.

\vskip1pc\noindent At this point it is also convenient to replace
the components of the stress tensor $T^{ab}$ by the corresponding
scalar quantities representing its projections  along the
directions tangential to the sphere and along its radius, ${\bf t}$ and ${\bf u}$: $T_\|=
t_a t_b T^{ab}$, $T_{\|\perp}= t_a u_b T^{ab}$, and $T_\perp= u_a
u_b T^{ab}$.\footnote{\sf With respect to the adapted coordinate
system, $t^a= (0,r^{-1})$ and $u^a=(1,0)$ so that $t_a=(0,r)$ and
$u_a=(1,0)$.} In particular, the tangential stress
\begin{equation}
T_\| = t_s t_s T^{ss} = r^{2} T^{ss}=  -  C_{\|} (s)\,{1\over r^2}
\label{eq:Tpar}
\end{equation}
diverges as $r^{-2}$. We see immediately that the consistency of
the shape equation with a conical geometry places a very strong
constraint on the admissible form of $T_{\|}$, determining
completely its radial dependence.

\vskip1pc\noindent Let us now examine the conservation law for
$T^{ab}$ to see what it tells us about $T_{\|}$, $T_\perp$ and
$T_{\|\perp}$. The tangential and radial projections of Eq.~(\ref
{eq:Tcons0}) read:
\begin{eqnarray}
 {1\over r}\,T_\| ' + {\partial \over \partial r} T_{\|\perp} + {2\over r} \, T_{\|\perp}
 &=& 0\,;\label{eq:T1}\\
 {\partial T_\perp\over \partial r} +  {1\over r}\, (T_\perp - T_\|) +
{1\over r} \, T_{\|\perp}'
 &=&0
 \,. \label{eq:T2}
\end{eqnarray}
Often the symmetry of the problem will imply $T_{\|\perp}=0$ so
that, modulo Eq.~(\ref{eq:T1}), $C_\|$ is a constant independent
of $s$. If the circular frame discussed in \cite{PRS} is replaced by one that is irregular,
however, the balance of forces on the frame will require
$T_{\|\perp}\ne0$.

\vskip1pc\noindent Remarkably, when $C_{\|}$ is constant, as it will if the disc is circular,
the shape equation assumes exactly the same form as the Euler-Lagrange
equation for planar Euler Elastica with tension $\sigma\propto
C_\|-1$ and $\kappa$ in place of the Frenet curvature, $\kappa_F$.

\vskip1pc\noindent The Euler-Lagrange equation for planar Euler
Elastica is completely integrable for $\kappa$ allowing the cone
to be reconstructed using the structure equations. The tension in
this fictitious Elastica can be either positive or negative,
depending on the boundary conditions. This should be compared to
the results of Cerda and Mahadevan who do not admit negative
values \cite{PRS}. Note that if $\kappa_F$ is used in place of
$\kappa$ the equation is different. The shape does not correspond
to Euler Elastica on a sphere \cite{LangerSinger}.

\vskip1pc\noindent More generally, the off-diagonal constraining stress is given by
\begin{equation}
T_{\|\perp} = C_\|(s)' \,{\ln r \over r^2}  + {C_{\|\perp}(s)\over
r^2}\,, \label{eq:Tparperp}
\end{equation}
where $C_{\|\perp}$ is some other function of $s$. In a moment, we
will see that if $C_{\|\perp}(s)$ is non-vanishing, it will play a
role in the balance of torques on the cone.

\vskip1pc\noindent The second conservation law (\ref{eq:T2})
identifies the radial scalar $T_\perp$ as a sum of five terms
\begin{equation}
T_\perp =  C_\| (s)''\,\left( {\ln r\over r^2 } +  {1\over r^2}
\right) + {C_\| (s) \over r^2} - {C_{\|\perp}(s)'\over r^2} +
{C_\perp(s)\over r}\,. \label{eq:Tperp}
\end{equation}
In particular, if $C_\|$ is constant and $C_{\|\perp}=0$,
\begin{equation} T_\perp = {C_{\|} \over r^2}  + {C_\perp (s)\over r}\,.
\end{equation}
The $r^{-2}$ behavior term is slaved to $T_\|$. This part of the
stress tensor $T^{ab}$ is traceless: $g^{ab} T_{ab}=0$, a property
that can be tracked back to the scale invariance of the bending
energy. The $r^{-1}$ behavior indicates the existence of a source
term associated with external forces.

\vskip1pc\noindent It is worth emphasizing  the following  points:
the stress associated with the constraint on the metric is
explicitly anisotropic and inhomogeneous. It is not necessarily
diagonal with respect to the orthonormal frame adapted to the
cone. While it is not monotonic, it does vanish asymptotically.

\vskip1pc\noindent  It is also useful to possess explicit
expressions for the surface components of the stress, ${\bf f}_\| =
t_a {\bf f}^a $ and ${\bf f}_\perp = u_a {\bf f}^a$ representing the
force per unit length transmitted across curves of constant $s$ and $r$
respectively \cite{MDG}. A straightforward calculation gives
\begin{eqnarray}
 {\bf f}_\| &=& {1\over r^2} \left[
\left( {1\over 2}\, \kappa^2 +
r^2 T_\|\right)\, {\bf t} + r^2 T_{\|\perp}\, {\bf u} - \kappa'\, {\bf n} \right] \label{eq:f||}
\\
{\bf f}_\perp &=& {1\over r^2} \left[ \left(- {1\over 2}\, \kappa^2
+ r^2 T_\perp\right)\, {\bf u} + r^2 T_{\|\perp} \, {\bf t} +
\kappa\, {\bf n} \right]\,.\label{eq:fperp}
\end{eqnarray}

\vskip1pc\noindent As a check of consistency, note that the divergence may be expressed as
\begin{equation}
\nabla_a {\bf f}^a = {1\over r} \, {\bf f}_\|' + {\partial {\bf
f}_\perp\over
\partial r} + {1\over r}\, {\bf f}_\perp
\label{eq:nabf}
 \end{equation}
In particular, it is simple to confirm that the projections of
Eq.~(\ref{eq:nabf}) along ${\bf n},{\bf t}$ and ${\bf u}$
reproduce Eqs.~(\ref{eq:ELkappaT}), (\ref{eq:T1}) and
(\ref{eq:T2}) respectively.


\section{\Large \sf Force and torque balance at constant $r$
\label{sec:forcetorque}}

\vskip1pc\noindent Now look at the balance of forces along a
closed curve $\Gamma_{r}$ of constant $r$. The physical arc-length
is $d\tau=rds$. The normal to this contour pointing away from the
apex is the vector ${\bf u}$. Then the total force transmitted
across the curve is given by\footnote{\sf Any closed curve
homotopic to $\Gamma_r$ can be used to determine the force
replacing ${\bf f}_\perp$ by $l^a {\bf f}_a$, where ${\bf l}=
l^a{\bf e}_a$ is the normal to the curve on the cone pointing
towards increasing $r$.}
\begin{eqnarray}
{\bf F} =  \oint_{\Gamma_{r}} d\tau \,{\bf f}_\perp  &=& {1\over r}
\oint ds\, \left( \kappa \, {\bf n} - {1\over 2} \kappa^2 \, {\bf u}
\right)  + r\, \oint ds \, T_\perp\, {\bf u} + r \oint ds\,
T_{\|\perp}\, {\bf t} \nonumber\\
&=&-{1\over r} \oint ds\, \left (1
+ {1\over 2} \kappa^2\right)\, {\bf u}+ r\, \oint ds \, T_\perp\,
{\bf u} + r \oint ds\, T_{\|\perp}\, {\bf t}
 \,, \label{eq:intf2}
\end{eqnarray}
where we have used Eq.~(\ref{eq:fperp}) on the
first line and Eq.~(\ref{eq:tpt}) on the second. A much more
transparent expression is possible in terms of the solution of the
conservation law Eq.~(\ref{eq:Tcons0}) for $T^{ab}$. Look at the
integral along any closed curve of the Euler-Lagrange derivative
${\cal E}$, weighted by ${\bf n}$. In equilibrium,
\begin{equation}
\oint ds \,\left(\kappa '' + {1\over 2} \, \kappa^3 + \kappa +
\kappa r^ 2T_\| \right) \,{\bf n} =0 \,.
\end{equation}
Now, performing an integration by parts on the first term and using
Eq.~(\ref{eq:tpn}), we obtain
\begin{equation}
\oint ds \,\kappa ''\, {\bf n} = - \oint ds \, \kappa \kappa' {\bf
t}\,.
\end{equation}
Furthermore, using Eq.~(\ref{eq:tpt}) to express ${\bf n}$ as a
linear combination of ${\bf u}$ and ${\bf t}'$, we obtain
\begin{eqnarray}
\oint ds \,\left({1\over 2} \, \kappa^3 + \kappa + \kappa r^2 T_\|
\right) \,{\bf n} &=& - \oint ds \,\left({1\over 2} \, \kappa^2 +
1
+ r^2  T_\| \right) \, ({\bf t}' + {\bf u})\nonumber\\
&=& \oint ds  \, \kappa \kappa' {\bf t}  + r^2 \oint ds \, T_\|'\,
{\bf t}  \nonumber\\
&& \quad - \oint ds \,\left({1\over 2} \, \kappa^2 + 1 + r^2 T_\|
\right) \, {\bf u}
 \,.
\end{eqnarray}
Thus, the tangential stress is constrained as follows on a closed cone:
\begin{equation}
\oint ds\,  \left (1 + {1\over 2} \kappa^2\right)\, {\bf u} + r^2
\oint ds\, T_\| \,{\bf u} - r^2 \oint ds \, T_\|'\, {\bf t} =0 \,.
\label{eq:ident0}
\end{equation}
Modulo this identity we can finally write the total force in terms
of the function $C_\perp$ appearing in Eq.~(\ref {eq:Tperp}) as
\begin{equation}
{\bf F} =  r\, \oint_{\Gamma_{r}} ds \, \left[ (T_\perp + T_\|)
{\bf u} + (T_{\|\perp} - T_{\|}' ) \, {\bf t} \right] =
\oint_{\Gamma_{r}} ds \,C_\perp(s) \, {\bf u} \, ,
\label{eq:forcewithoutgravity}
\end{equation}
where we have used the expressions for
the projections of the stress tensor (\ref{eq:Tpar}),
(\ref{eq:Tparperp}) and (\ref{eq:Tperp}) consistent with the shape
equation and the conservation of $T^{ab}$. The line
integration does not depend on $r$. If an
external force is acting, the term proportional to $r^{-1}$ in
$T_{\perp}$ does not vanish. If a force ${\bf F}$ is applied at
the apex of a cone supported on a frame,
$C_\perp(s)$ is the distribution of forces
on the rim of this frame needed to counterbalance it. Notice that the tangential and the
radial projections of the conservation law can be written as
\begin{equation}
\nabla_a T^{ab}= \left[ C_\perp (s)\, u^b + C_{\|\perp} (s)
t^b\right] \,  {\delta(r)\over r} \,. \label{eq:nTdel}
\end{equation}

\vskip1pc\noindent In exactly the same way, we can consider the
torque. The conserved torque tensor is given by \cite{stress,balancingtorques}
\begin{equation}
  {\bf m}^a = {\bf X} \times {\bf f}^a + K g^{ab} {\bf e}_{b}\times {\bf n}
  \label{eq:torquetensor}
\end{equation}
and the total torque about the origin can be written as:
\begin{equation}
{\bf M} = \oint_{\Gamma_{r}} \!\!\romd \tau\; u_{a}{\bf m}^a \,.
\end{equation}
Thus
\begin{eqnarray} {\bf M} &=& r \oint_{\Gamma_{r}} ds \,
\Big[\big( r {\bf u} \times {\bf f}_\perp \big) + K {\bf u} \times {\bf
n} \Big] \nonumber\\
&=& -2 \,\oint_{\Gamma_{r}} ds \,\kappa {\bf t}
+ r^2 \oint_{\Gamma_{r}} ds \, T_{\|\perp} \, {\bf n} \nonumber\\
&=& \oint_{\Gamma_{r}} ds\, C_{\|\perp}(s)\, {\bf n}\,.
\label{eq:totaltorque}
\end{eqnarray}
The integrated contribution from ${\bf f}^a_0$ vanishes on account
of the structure Eq.~(\ref{eq:tpn}) and the closure of
$\Gamma_r$.
Furthermore, the term proportional to $\ln{r}/r^{2}$ in
$T_{\|\perp}$ [see Eq.~(\ref{eq:Tparperp})] does not contribute to
the torque as one can show by integrating the Euler-Lagrange
derivative~(\ref{eq:ELkappaT}) weighted by $\VECt$ along a closed curve.

\vskip1pc\noindent Eq.~(\ref{eq:totaltorque}) implies that the cone
geometry may provide an equilibrium when an external torque is
applied. In that case, the off-diagonal stresses do not vanish and
the value of the torque is captured by $C_{\|\perp}(s)$.


\section{\Large \sf Brought down by gravity\label{sec:gravity}}

Suppose that the sheet is held at a point and allowed to fall
under gravity. The bending energy is minimized by the planar disc;
gravitational potential energy is minimized by a sheet that hangs
vertically; as described in \cite{drape}, the competition between
the two gives rise to draping patterns.

\vskip1pc\noindent If the weight of the sheet is taken into account,
its gravitational potential energy is given by
\begin{equation} H_g[{\bf X}] = -\rho g \int dA\,
{\bf X}\cdot {\bf z}\,,
\end{equation}
where $\rho$ is the density of the sheet and $g$ is the acceleration due
of gravity. To counterbalance the weight of the sheet an external
force $\VECF_{\text{ext}}$ in the $\VECz$-direction has to be
applied (see Fig.~\ref{fig:cone}). This force acts at the apex of
the cone (see previous section).

\vskip1pc\noindent Under the deformation, ${\bf X}\to {\bf X} +
\delta {\bf X}$, the variation of $H_g$ is given by
\begin{equation}
 \delta H_g = - \rho g\, \int dA\,
( K {\bf X}\cdot {\bf z} + {\bf n}\cdot {\bf z} )\, {\bf n}\cdot
\delta {\bf X}\,.
\end{equation}
The effect is to add a normal source to the conservation law~(\ref{eq:fcons}):
\begin{eqnarray}
\nabla_a {\bf f}^a &=& \rho g \,( K {\bf X}\cdot {\bf z} + {\bf
n}\cdot {\bf z} )\, {\bf n}\nonumber\\
 &=& \rho g   (\kappa
{\bf u}\cdot {\bf z} + {\bf n}\cdot {\bf z} )\, {\bf n} = j (s)\,
{\bf n}\,.
\end{eqnarray}
The Euler-Lagrange equation~(\ref{eq:ELkappaT}) is replaced by
the equation \begin{equation}
- {1\over r^3} \left(\kappa '' + {1\over 2} \, \kappa^3 + \kappa
\right) - {\kappa\,\over r}  \, T_\|  = j (s)\,.
\end{equation}
This differs from the result that Cerda \etal have derived. They
average over the radial direction \cite{drape}.

\vskip1pc\noindent As before, consistency will place a constraint
on $T_{\|}$. However, we now require that\footnote{\sf In the following,
all variables with an upper index ${\sf g}\!=\!0$ refer to the case
without gravity.}
\begin{equation}
  T_{\|} = - {C_\|(s)\over r^2 } - {j \over \kappa} \, r
  = T_{\|}^{{\sf g}=0} - {j \over \kappa} \, r
  \; .
  \label{eq:Tparg}
\end{equation}
Even in the regime where the analysis provided in appendix~\ref{app:auxvariables}
is valid, with gravity we begin to see how such an approach breaks
down. The pre-averaging over $r$ provides a different (and
incorrect) shape equation.

\vskip1pc\noindent The conservation law~(\ref{eq:Tcons0}) for $T^{ab}$ is unchanged
as gravity only contributes a normal source. However, because of Eq.~(\ref{eq:Tparg}),
in place of Eqs.~(\ref{eq:Tparperp}) and (\ref{eq:Tperp}) we
have
\begin{equation}
T_{\|\perp} = T_{\|\perp}^{{\sf g}=0}  + \left({j\over
\kappa}\right)'\, {r\over 3} \,, \label{eq:Tparperpg}
\end{equation}
and
\begin{equation}
T_\perp =  T_\perp^{{\sf g}=0} - \left[{ j\over\kappa}\, +
{1\over 3}\, \left({j\over\kappa}\right)''\right] \, {r\over 2}\,.
\label{eq:Tperpg}
\end{equation}
The force transmitted across a curve of constant $r$ is then given
by
\begin{eqnarray}
  \VECF & = & \oint_{\Gamma_{r}} d\tau \,{\bf f}_\perp
  = r \oint_{\Gamma_{r}} ds \, \left[ (T_\perp + T_\|^{\text{g=0}}) {\bf u} +
    (T_{\|\perp} - T_\|'^{\;\text{g=0}}) \, {\bf t} \right]
  \nonumber \\
  & \stackrel{(\ref{eq:forcewithoutgravity})}{=} & \oint_{\Gamma_{r}} ds \,C_\perp(s) \, {\bf u}
    + r^{2} \oint_{\Gamma_{r}} ds \, \left\{ \frac{1}{2}\Big[{j\over\kappa}\, +
    {1\over 3}\, \left({j\over\kappa}\right)''\Big] {(-\bf u)} + \frac{1}{3}\left({j\over
    \kappa}\right)' \, {\bf t} \right\}
  \nonumber \\
  & \stackrel{(\ref{eq:tpt})}{=} &
    \oint_{\Gamma_{r}} ds \,C_\perp(s) \, {\bf u} + {r^2\over 2} \oint_{\Gamma_{r}} ds \, j\, {\bf n}
    -\frac{r^{2}}{6} \oint_{\Gamma_{r}} ds \, \left[\left({j\over\kappa}\right)' {\bf u}\right]'
  = - \VECF_{\text{ext}} + \int_{\Sigma_{r}} \romd A \; j \VECn
  \,, \;\;\;\;\;\;\;
\end{eqnarray}
where $\Sigma_{r}$ is that part of the surface which is enclosed by
the curve $\Gamma_{r}$. Note that the line integration does now depend on
$r$ but for good reason: the surface \emph{below} the curve $\Gamma_{r}$ exerts the
force $\VECF$ on $\Gamma_{r}$ due to gravity. Its absolute value
decreases quadratically for increasing $r$ and vanishes for $r=R$ as
expected.

\vskip1pc\noindent The torque about the origin is now given by
\begin{equation}
  \VECM = r^{2} \oint_{\Gamma_{r}} ds \, T_{\|\perp} \VECn
  \stackrel{(\ref{eq:Tparperpg})}{=}
  = \VECM^{\text{g=0}} + r^2 \oint_{\Gamma_{r}} ds \, \left({j\over
  \kappa}\right)'\, {r\over 3} \,\VECn
  = \VECM^{\text{g=0}} - \int_{\Sigma_{r}} dA \; r \, j\,\VECt
  \; .
\end{equation}
The last term due to gravity can also be written as
$\int_{\Sigma_{r}} dA \, [\VECX\times(j\VECn)]$, again exactly as
one would expect.


\section{\Large \sf Discussion}

In this paper we have introduced a geometrical framework to
examine the bending of an (unstretchable) sheet of paper. We have
illustrated the viability of this framework by examining the
conical deformations of a planar sheet. In particular, we have
examined in some detail the distribution of stresses within the
sheet associated both with bending and with the constraint and how
they conspire to transmit the external forces and torques acting
on the sheet.

\vskip1pc\noindent Various characteristic patterns are associated
with this geometry: the radial dependence of the stress is
constrained strongly by the geometry; how it varies along the
spherical generating curves will depend on the specific boundary
conditions associated with the external forces acting on the
sheet. We have also shown how the presence of bulk forces such as
gravity are accommodated within this framework. In the latter
case, we point out the pitfalls of prematurely constraining the
geometry within the variational principle.

\vskip1pc\noindent We have not attempted to discuss the physical
details within the ridges and peaks where the elastic
approximation breaks down. Various elements of our framework are,
however, likely to play a role in any refinement of the model
which does address this physics.

\vskip1pc\noindent We have limited our discussion of the
application of our framework to conical shapes; this is because
they are relatively simple not because they are the only
interesting configurations. It would be interesting (and it should
be straightforward) to examine the M{\"o}bius strip geometry
described recently by Starostin and van der Heijden within our
framework \cite{star}.

\vskip1pc\noindent It is appropriate to think of a cone as a kind
of elementary deformation of a planar sheet in which the energy is
localized. A generic patch of flat surface, however, will be
described by a tangent developable surface. Such a surface is
singular along a certain curve: its edge of regression which will
itself generally exhibit singularities \cite{Struik,Gray}. The
flat directions are generated by the tangent to this curve.
Typically, we do not see this curve when we fold paper. The patch
gets truncated by a ridge or we run off the sheet before the
corresponding generating curve is reached. However, the folded
sheet is, in principal, completely described by this set of
curves. This description will be the subject of a subsequent
publication \cite{Pablo}.

\vskip3pc \noindent{\large \sf Acknowledgments}

\vspace{.5cm}

\vskip1pc\noindent  We thank Riccardo Capovilla for getting us interested in this problem.
We have benefitted from conversations with him and Markus Deserno. Timo Sch{\"u}rg gave us
some valuable tips on how to use POV-Ray. Partial support from
CONACyT grant 51111 as well as DGAPA PAPIIT grant IN119206-3 is
acknowledged.


\appendix

\section{\sf \Large Fixing Gaussian curvature \label{app:Gaussiancurvature}}

\vskip1pc\noindent Let us examine the consequences of fixing the
Gaussian curvature instead of the metric.  We thus construct the
constrained functional
\begin{equation}
H_C[{\bf X}]= H[{\bf X}] + \int dA\, W(u^a)\, K_{\romG} \,,
\end{equation}
where $K_{\romG}$ is the Gaussian curvature \cite{notation}. Here
$W(u^a)$ is a local Lagrange multiplier constraining $K_{\romG}$
to vanish. Analogous functionals can be defined for a constraint
on any other geometrical scalar.

 \vskip1pc\noindent The
constrained equilibrium may again be expressed as a conservation law
of the form (\ref{eq:fcons}). The stress tensor ${\bf
f}^a$ is again given by an expression of the form
(\ref{eq:stressdef}) with an additional tangential stress $T^{ab}$.
However, this stress is now determined completely by the multiplier $W$,
\begin{equation}
T^{ab}= {1\over 2}\left(\nabla^a\nabla^b W- g^{ab} \nabla^2
W\right)\,. \label{eq:Tdef}
\end{equation}
The constraint again introduces tension in the surface but, this time, with a
single degree of freedom.

\vskip1pc\noindent It is appropriate to point out that there
is an ambiguity inherent in the definition of the stress tensor
${\bf f}^a$.  This ambiguity is due, in part, to the Gauss-Codazzi
integrability condition which connects the intrinsic and extrinsic
geometry \cite{Spivak}. If the variations are treated
intrinsically, one obtains the expression we have written down for
the stress; if, however, they are treated extrinsically an
apparently very different answer is obtained. The two are entirely
consistent; they differ only by a conserved null stress that does
not transmit forces. The relevant calculations have appeared in a
different context (see, for example, \cite{auxil}).

\vskip1pc\noindent Whereas the conservation of $T^{ab}$ is an important constraint
when we fix the metric, here it is simple to confirm that $T^{ab}$
is automatically conserved if the geometry is flat.
This is a consequence of the definition of $T^{ab}$ in
Eq.~(\ref{eq:Tdef}) in terms of a potential.
In analogy with elasticity theory, it is appropriate to think of
$W$ as an Airy potential for $T^{ab}$ \cite{LL}.

\vskip1pc\noindent Note that the constraints break the invariance
of the two dimensional bending energy with respect to conformal
transformations of three-dimensional space. In general, the
Gaussian curvature is not preserved. It is simple to show that
under inversion in the origin ${\bf X}\to {\bf X}/|{\bf X}|^2$,
one has \cite{Gray}
\begin{equation}
 K_{\romG} \to    |{\bf X}|^4\,
K_{\romG} - \,2 \, ({\bf X}\cdot {\bf n})|{\bf X}|^2\, K   + \,4
\, ({\bf X}\cdot {\bf n})^2\,. \label{eq:KGKG}
\end{equation}
Remarkably, the only geometries which remain flat are the cones we
considered with apex at the origin, and which thus satisfy ${\bf
X}\cdot {\bf n}=0$. Given one cone, its inversion in the origin is
another \cite{Pavel}. Note, however, that curves on the unit sphere
are invariant under inversion in the origin. Thus, while the physics might be different,
the cone remains the same.


\section{\Large \sf Cones as trajectories on spheres \label{app:auxvariables}}


\subsection{\Large \sf Derivation of the Euler-Lagrange equations\label{subsec:ELEaux}}

\vskip1pc\noindent
If we introduce the
cutoff $r_0$ at the apex and integrate over the radial direction,
the bending energy of a cone formed by a circular disc with its apex at the centre
is given by
\begin{equation}
H [{\bf u}]  = \frac{1}{2}\, a \, \oint_{\Gamma} ds \, \kappa
^2\,, \label{eq:HS2}
\end{equation}
where $a$ is given by
\begin{equation}
  a= \ln (R/r_0)
  \, .
\end{equation}
The dimensional dependence is contained completely within this
logarithm. This behavior is a consequence of the scale invariance
of $H$. $H$ is a functional of curves on the unit sphere.

\vskip1pc\noindent We have already pointed out in
section~\ref{subsec:coneconstrained} that if the boundary
conditions are not symmetrical, the identification with Euler
Elastica breaks down. It is, however, useful to see how the shape
equation emerges from a variational principle which exploits the
mapping from cones into trajectories on a sphere. Our task then is
to identify the trajectories on a sphere minimizing the
Hamiltonian $H$ given by Eq.~(\ref{eq:HS2}).

\vskip1pc\noindent Although the Euler-Lagrange equation itself is
simple, its derivation is quite subtle. As we have found
elsewhere, it is useful to introduce an appropriate set of
auxiliary variables associated with local geometrical constraints
\cite{auxil}. Consider the functional
\begin{equation}
F [{\bf u} , {\bf t} , \lambda , \Lambda , {\bf f} ] = \oint_{\Gamma}
ds \; \left[ \frac{a}{2} ( {\bf u} \cdot {\bf t} \times {\bf t}'
)^2 + \frac{\lambda}{2} ( {\bf u}^2 - 1 ) + \frac{\Lambda}{2} ( {\bf
t}^2 - 1 ) + {\bf f} \cdot ({\bf t} - {\bf u}' ) \right]\,,
\label{eq:Hc}
\end{equation}
where $\lambda(s)$ and $\Lambda(s)$ are two local Lagrange multipliers
enforcing the constraints that ${\bf u}$ is a unit vector and that
the parameter $s$ is arc-length on the sphere. The area constraint
is thus implemented by fixing the range of integration on $s$ to
the interval $[0,2\pi]$.

\vskip1pc\noindent Variation of $F$ with respect to ${\bf u}$
gives
\begin{equation}
{\bf f}' = - (a \kappa^2 + \lambda)\, {\bf u} + a \kappa \,{\bf n}
\,, \label{eq:fp1}
\end{equation}
The right hand side of  Eq.~(\ref{eq:fp1})  has been simplified by
using the structure equations~(\ref{eq:tp}) for the curve. Note,
in particular, that ${\bf f}$ is not constant as it would be for a
translationally invariant Hamiltonian. In the functional $F$
translational invariance is broken both in the energy itself and
in the constraint on ${\bf u}$.

\vskip1pc\noindent Variation with respect to ${\bf t}$ identifies
the vector ${\bf f}$ appearing in Eq.~(\ref{eq:fp1}) as
\begin{equation}
{\bf f} = -  (\Lambda + 2a\kappa^2) \, {\bf t} - a \kappa' {\bf n} \;
, \label{eq:ftn1}
\end{equation}
Later, we will provide a physical interpretation of ${\bf f}$ as
an effective force per unit length along a ray of fixed $s$.

\vskip1pc\noindent  By substituting the expression for ${\bf f}$
given by Eq.~(\ref{eq:ftn1}) into Eq.~(\ref{eq:fp1}) we obtain three
equations describing the equilibrium. The projections along ${\bf
u}$ and along ${\bf t}$ determine the multipliers $\lambda$ and
$\Lambda$. The remaining equation along ${\bf n}$ is the
Euler-Lagrange equation.

\vskip1pc\noindent Note that
\begin{equation}
{\bf f} \cdot {\bf u} =0\,. \label{eq:fx}
\end{equation}

\vskip1pc\noindent Let us first examine the component
of the conservation law along ${\bf t}$. We note that
\begin{equation}
 {\bf f}'\cdot {\bf t} \stackrel{(\ref{eq:fp1})}{=}0\,\,.
\label{eq:fpt}
\end{equation}
This equation fixes the function  $\Lambda$ appearing in
Eq.~(\ref{eq:ftn1}) up to a constant. We have
\begin{equation}
\Lambda = c - \frac{5 a}{2} \kappa^2 \,,
\end{equation}
where $c$ is a constant. We thus identify
\begin{equation}
{\bf f} =  \left(\frac{1}{2} a\kappa^2 - c\right) \, {\bf t} - a
\kappa' {\bf n} \,. \label{eq:fc}
\end{equation}
Comparing the two stress tensor projections (\ref{eq:f||}) and
(\ref{eq:fperp}) from Sec.~\ref{subsec:coneconstrained} with the
Lagrange multiplier $\VECf$ yields the remarkably simple relation
$\VECf=ar^{2}\VECf_{\|}$ if $T_{\|\perp}=0$.
This confirms that, when this approximation is valid, $\VECf$ is
the tangent stress along a ray of fixed $s$.

\vskip1pc\noindent To determine $\lambda$, note that
\begin{equation}{\bf f}' \cdot {\bf u} \stackrel{(\ref{eq:fp1})}{=}
-( a \kappa^2 + \lambda )\,.
\end{equation}
Using Eq.~(\ref{eq:fc}), it follows that
\begin{equation}
\lambda  = - \frac{a}{2} \kappa^2 - c \,.
\end{equation}
Note that this variable does not appear in the final shape
equation which involves normal projections. For the record, we
have
\begin{equation}
{\bf f}'=-\left(\frac{a}{2}\kappa^2 -c\right) \, {\bf u} +
a\kappa\, {\bf n}\,. \label{eq:fpp}
\end{equation}
Its normal projection together with Eq.~(\ref{eq:fc}) provides the
Euler-Lagrange equation
\begin{equation}
-a\,  \left( \,\kappa'' + \frac{1}{2}\,\kappa^3 + \kappa\right) +
c \, \kappa  =0 \,. \label{eq:Elk}
\end{equation}
The constant $c$ is a tension associated with the fixed area
constraint which is implied by fixing the arc-length. If this
constraint is relaxed, $c=0$; however, there persists a `tension'
proportional to $a$ associated with constraining the curve
trajectory to a sphere. Earlier we identified the origin of this
term in the $r$ dependence of $\nabla^2 K$ on the cone (see
Sec.~\ref{subsec:coneconstrained}). We reproduce the shape
equation (\ref{eq:Elk}) obtained earlier if we identify $C_\|$
with the ratio of the constants ($c/a$) (when $C_\|$ is constant).

\vskip1pc\noindent Generally, however, it appears to be beyond the
scope of the variational analysis based on the reduced Hamiltonian
to provide a consistent description of the cone. In particular,
it does not appear to be possible to accommodate $T_{\|\perp}\ne
0$. Even when $T_{\|\perp} =0$, there is no obvious  way, within
this simplified framework, to analyze the $r$ dependent stresses
set up in the cone as they, for instance, show up in the presence
of gravity (see section~\ref{sec:gravity}).

\vskip1pc\noindent We note that there is an integrability
condition associated with Eq.~(\ref{eq:fpp}). The identity $\oint
ds\, {\bf f}'=0$ associated with closure implies that the
integrated source of ${\bf f}$ must also vanish. Thus
\begin{equation}
\oint ds\, \Big[\left(\kappa^2 - 2 c/a\right) \, {\bf u} -2
\kappa\, {\bf n}\Big] =0 \,.\label{eq:fppp}
\end{equation}
This identity can be simplified further using Eq.~(\ref{eq:tpt})
to give
\begin{equation}
\oint ds\, \left(\kappa^2 - 2 \sigma/a\right) \, {\bf u}  =0 \,.
\label{eq:fp4}
\end{equation}
There is no obvious physical interpretation of this identity
within this one-dimensional framework. It is instructive to
compare Eq.~(\ref{eq:fp4}) with its analogue Eq.~(\ref{eq:ident0})
within the general framework which we used to place a constraint
on the stresses set up in the cone.


\subsection{\Large \sf Rotations and Conservation
laws\label{subsec:conservationlaws}}

\vskip1pc\noindent Fixing the apex of the cone at the origin
breaks translational invariance. The residual symmetry is
three-dimensional rotational invariance. The shape equation
describing the sheet may be identified as the conservation law
associated with this symmetry. In equilibrium, we must have
\begin{equation}
\delta F= - \int ds \frac{d}{ds} [  {\bf f}\cdot \delta {\bf u} +
a \kappa ({\bf u}\times {\bf t}) \cdot \delta {\bf t} ] \,.
\end{equation}
In particular, $\delta F=0$ under rotation. With $\delta {\bf u}=
{\bf b}\times {\bf u}$ and $\delta {\bf t}={\bf b}\times {\bf t}$,
we have
\begin{equation}
 {\bf b}\cdot  [{\bf u}\times {\bf f}  + a \kappa \, {\bf u}]
\end{equation}
is constant.  Thus
\begin{equation}
{\bf u}\times {\bf f} + a \kappa \, {\bf u} = {\bf J}\,,
\label{eq:XF}
\end{equation}
where ${\bf J}$ is a constant vector which can be identified as the
tangential projection of the torque tensor~(\ref{eq:torquetensor})
\begin{equation}
  t_{a}\VECm^{a} = \VECX \times \VECf_{\|} + K\VECu
    = \VECX\times \frac{\VECf}{a r^2} + \frac{\kappa}{r}\VECu
    = \frac{1}{a r} \VECJ
  \; ,
\end{equation}
if $T_{\|\perp}=0$.

\vskip1pc\noindent Using the structure equations~(\ref{eq:tp})
and the Euler-Lagrange
equation~(\ref{eq:Elk}) one can easily show that ${\bf J}$ is
indeed conserved. Squaring, using the identify ${\bf f}\cdot {\bf
u}=0$, we find
\begin{equation}
{\bf f}^2 + a^2 \kappa^2 = {\bf J}^2\,.
\end{equation}
This is the first integral of the Euler-Lagrange equation.
It can be rewritten
\begin{equation}
  (\kappa')^{2} + \frac{1}{4}\kappa^{4} - \frac{\sigma}{a}\kappa^{2}
  =\frac{{\bf J}^2 - c^{2}}{a^{2}}
  \; ,
\end{equation}
and integrated after a separation of variables.


\end{document}